\documentclass[a4paper,12pt]{article}
\usepackage{amsmath,amssymb,amsthm}

\newtheorem{lemma}{Lemma}
\newtheorem{theorem}{Theorem}

\theoremstyle{definition}

\theoremstyle{remark}

\newcommand{\beq}{\begin{eqnarray}}
\newcommand{\eeq}{\end{eqnarray}}
\newcommand{\beqnn}{\begin{eqnarray*}}
\newcommand{\eeqnn}{\end{eqnarray*}}
\newcommand{\rd}{\partial}

\newcommand{\CC}{\mathbf{C}}
\newcommand{\PP}{\mathbf{P}}

\newcommand{\ZZ}{\mathbf{Z}}
\newcommand{\bst}{\boldsymbol{t}}
\newcommand{\bsT}{\boldsymbol{T}}
\newcommand{\bsx}{\boldsymbol{x}}
\newcommand{\bsy}{\boldsymbol{y}}

\newcommand{\Bbar}{\bar{B}}
\newcommand{\ubar}{\bar{u}}
\newcommand{\Lbar}{\bar{L}}
\newcommand{\Tbar}{\bar{T}}
\newcommand{\bsTbar}{\bar{\bsT}}

\begin{document}

\title{Integrable structure of melting crystal model\\
with two $q$-parameters}
\author{Kanehisa Takasaki%
\thanks{E-mail: takasaki@math.h.kyoto-u.ac.jp}\\
{\small
Graduate School of Human and Environmental Studies,
Kyoto University}\\
{\small Yoshida, Sakyo, Kyoto, 606-8501, Japan}}
\date{}
\maketitle

\begin{abstract}
This paper explores integrable structures of a generalized 
melting crystal model that has two $q$-parameters $q_1,q_2$.  
This model, like the ordinary one with a single $q$-parameter, 
is formulated as a model of random plane partitions (or, 
equivalently, random 3D Young diagrams).  The Boltzmann weight 
contains an infinite number of external potentials that depend 
on the shape of the diagonal slice of plane partitions.  
The partition function is thereby a function of an infinite number 
of coupling constants $t_1,t_2,\ldots$ and an extra one $Q$. 
There is a compact expression of this partition function 
in the language of a 2D complex free fermion system, from which 
one can see the presence of a quantum torus algebra behind 
this model.  The partition function turns out to be 
a tau function (times a simple factor) of two integrable 
structures simultaneously.  The first integrable structure 
is the bigraded Toda hierarchy, which determine the dependence 
on $t_1,t_2,\ldots$.  This integrable structure emerges 
when the $q$-parameters $q_1,q_2$ take special values.  
The second integrable structure is a $q$-difference analogue 
of the 1D Toda equation.  The partition function satisfies 
this $q$-difference equation with respect to $Q$.  Unlike 
the bigraded Toda hierarchy, this integrable structure 
exists for any values of $q_1,q_2$.  
\end{abstract}
\bigskip

\begin{flushleft}
\end{flushleft}
\newpage

\section{Introduction}

The {\it melting crystal model} is a model of 
statistical physics and describes a melting corner 
of a crystal that fills the first quadrant of 
the 3D Euclidean space.  The complement of 
the crystal in the first quadrant may be 
thought of as a 3D analogue of Young diagrams. 
These 3D Young diagram can be represented 
by {\it plane partitions}.  
Thus the melting crystal model can be formulated 
as a model of {\it random plane partitions}.  

This model has been applied to string theory 
\cite{ORV03} and gauge theory \cite{MNTT04a,MNTT04b}.  
From the point of view of gauge theory, 
the partition function of the melting crystal model 
is a 5D analogue of the instanton sum 
of 4D $\mathcal{N}=2$ supersymmetric 
Yang-Mills theory \cite{Nekrasov02,NY05,NO03}.  
(Curiously, the 4D instanton sum also resembles 
a generating function the Gromov-Witten invariants 
of the Riemann sphere \cite{OP02a,OP02b}.) 
This analogy will need further explanation, 
because the 4D instanton sum is a sum over 
ordinary partitions rather than plane partitions. 
The fact is that one can use the idea 
of {\it diagonal slicing} \cite{OR03} 
to rewrite the partition function of 
the melting crystal model to a sum 
over ordinary partitions \cite{MNTT04a}.  
Comparing these two models of {\it random partitions}, 
one can consider the melting crystal model 
as a kind of {\it $q$-deformation} of 
the 4D instanton sum. Here $q$ is a parameter 
of the melting crystal model related to 
temperature.  

In our previous work \cite{NT07} (see also 
the review \cite{Takasaki08}), we introduced 
a set of external potentials into this model, 
and identified an integrable structure 
that lies behind this partition function. 
Namely, the partition function, as a function 
of the coupling constants $t_1,t_2,\ldots$ 
of potentials, turns out to be equal to 
a tau function (times a simple factor) 
of the Toda hierarchy \cite{UT84,TT95}.  
Moreover, the tau function satisfy a set of 
constraints that reduces the full Toda hierarchy 
to the so called {\it 1D Toda hierarchy}.  
Though a similar fact was known for 
the 4D instanton sum \cite{Losev-etal03,MN06,Marshakov07}, 
we found that the partition function 
of the melting crystal model can be treated 
in a more direct manner.   We derived these results 
on the basis of a fermionic formula of 
the partition function \cite{Losev-etal03}.   
A technical clue is a set of algebraic relations 
among the basis of a quantum torus (or cylinder) algebra 
realized by fermions.  These relations enabled us 
to rewrite the partition function to a tau function 
of the Toda hierarchy. 

In the present paper, we generalize these results 
to a melting crystal model with two $q$-parameters 
$q_1,q_2$ \cite{IKS08}.  Actually, 
since the potentials have another $q$-parameter $q$, 
this model has altogether three $q$-parameters 
$q_1,q_2$ and $q$;  letting $q_1 = q_2 = q$, 
we can recover the previous model.  

Our goal is two-fold.  Firstly, we elucidate 
an integrable structure that emerges when 
$q_1$ and $q_2$ satisfy the relations 
$q_1 = q^{1/N_1}$ and $q_2 = q^{1/N_2}$ for 
a pair of positive integers $N_1$ and $N_2$. 
The partition function in this case turns out 
to be, up to a simple factor, a tau function of 
(a variant of) the {\it bigraded Toda hierarchy} 
of type $(N_1,N_2)$ \cite{Carlet06}, which is 
also a reduction of the Toda hierarchy. 
Secondly, without such condition on 
the parameters $q_1,q_1$ and $q$, 
we show that the partition function 
satisfies a $q$-difference analogue 
\cite{KS91,MMV94,AHvM98,Takasaki05}
of the Toda equation with respect to 
yet another coupling constant $Q$.  
In the gauge theoretical interpretation, 
$Q$ is related to the energy scale $\Lambda$ 
of supersymmetric Yang-Mills theory. 

This paper is organized as follows.  
Section 2 is a review of combinatorial aspects 
of the usual melting crystal model.  The model 
with two $q$-parameters is introduced in the end 
of this section.  Section 3 is an overview of 
the fermionic formula of the partition function.  
After reviewing these basic facts, 
we present our results on integrable structures 
in Sections 4 and 5.  
Section 4 deals with the bigraded Toda hierarchy, 
and Section 5 the $q$-difference Toda equation.   
We conclude this paper with Section 6.

\section{Melting crystal model}

In the following, we shall use a number of 
notions and results on partitions, Young diagrams 
and Schur functions.  For details of those 
combinatorial tools, we refer the reader 
to Macdonald's book \cite{Macdonald-book}.  
See also Bressoud's book \cite{Bressoud-book} 
for related issues and historical backgrounds.

\subsection{Simplest model}

Let us start with a review of the ordinary 
melting crystal model with a single parameter 
$q$ ($0 < q <1$).   As a model of statistical physics, 
this system can take various states with some probabilities, 
and these states are represented by plane partitions. 

Plane partitions are 2D analogues of ordinary 
(one-dimensional) partitions $\lambda 
= (\lambda_1,\lambda_2,\ldots)$, and denoted by 
2D arrays 
\beqnn
\pi = (\pi_{ij})_{i,j=1}^\infty
= \left(\begin{array}{ccc}
  \pi_{11} & \pi_{12} & \cdots \\
  \pi_{21} & \pi_{22} & \cdots \\
  \vdots   & \vdots   & \ddots 
  \end{array}\right)
\eeqnn
of nonnegative integers $\pi_{ij}$ 
(called {\it parts}) such that 
only a finite number of parts are non-zero 
and the inequalities 
\beqnn
  \pi_{ij} \ge \pi_{i,j+1}, \quad 
  \pi_{ij} \ge \pi_{i+1,j} 
\eeqnn
are satisfied.  Let $|\pi|$ denote the sum 
\beqnn
  |\pi| = \sum_{i,j=0}^\infty \pi_{ij}
\eeqnn
of all parts $\pi_{ij}$.  

Such a plane partition $\pi$ represents 
a 3D Young diagram in the first quadrant 
$x,y,z \ge 0$ of the $(x,y,z)$ space. 
In this geometric interpretation, $\pi_{ij}$ 
is equal to the height of the stack of cubes 
over the $(i,j)$-th position of the base $(x,y)$ plane. 
Therefore $|\pi|$  is equal to the volume of 
the 3D Young diagram.   

In the formulation of the melting crystal model, 
the complement of the 3D Young diagram 
in the first quadrant embodies the shape of 
a partially melted crystal.  We assume that 
such a crystal has energy proportional to $|\pi|$. 
Consequently, the partition function of this system 
is given by the sum 
\beqnn
  Z = \sum_{\pi} q^{|\pi|} 
\eeqnn
of the Boltzmann weight $q^{|\pi|}$ 
over all plane partitions $\pi$.

\subsection{Diagonal slicing}

We can convert this model of random plane partitions 
to a model of random partitions by diagonal slicing. 
This idea originates in the work of 
Okounkov and Reshetikhin \cite{OR03} 
on a model of stochastic process (Schur process).  

Let $\pi(m)$ ($m \in \ZZ$) denotes the partition 
that represents the Young diagram obtained 
by slicing the 3D Young diagram along 
the diagonal plane $x - y = m$ in the $(x,y,z)$ space.  
In terms of the parts $\pi_{ij}$ of the plane partition, 
these diagonal slices can be defined as 
\beqnn
  \pi(m) = 
  \begin{cases}
    (\pi_{i,i+m})_{i=1}^\infty &\mbox{if}\quad m \ge 0\\
    (\pi_{j-m,j})_{j=1}^\infty &\mbox{if}\quad m < 0 
  \end{cases}
\eeqnn
The plane partition can be recovered from 
this sequence $\{\pi(m)\}_{m=-\infty}^\infty$ of partitions.  
To be diagonal slices of a plane partition, 
however, these partitions cannot be arbitrary, 
and have to satisfy the condition 
\beq
  \cdots \prec \pi(-2) \prec \pi(-1) \prec \pi(0) 
  \succ \pi(1) \succ \pi(2) \succ \cdots. 
\eeq
Here ``$\succ$'' denotes the {\it interlacing relation} 
\beqnn
  \lambda = (\lambda_1,\lambda_2,\ldots) 
  \succ \mu = (\mu_1,\mu_2,\ldots) 
  \;\Longleftrightarrow\; 
  \lambda_1 \ge \mu_1 \ge \lambda_2 \ge \mu_2 \ge \cdots, 
\eeqnn
namely, $\lambda \succ \mu$ means 
that the skew diagram $\lambda/\mu$ 
is a {\it horizontal strip}.  

These diagonal slices, in turn, determine 
two Young tableaux 
\beqnn
  T = (T_{ij})_{(i,j)\in\lambda}, \quad 
  T' = (T'_{ij})_{(i,j)\in\lambda}
\eeqnn
on the main diagonal slice 
\beqnn
  \lambda = \pi(0)
\eeqnn
as 
\beqnn
\begin{aligned}
  T_{ij} = m \quad &\mbox{if}\quad (i,j) \in \pi(-m)/\pi(-m-1)\\
  T'_{ij} = m \quad &\mbox{if}\quad (i,j) \in \pi(m)/\pi(m+1). 
\end{aligned}
\eeqnn
Since the skew diagrams $\pi(\pm m)/\pi(\pm(m+1))$ 
are horizontal strips, these Young tableaux 
turn out to be {\it semi-standard tableaux}, 
namely, satisfy the inequalities%
\footnote{Note that the entries of the tableaux 
are arrayed in decreasing order rather than 
usual increasing order.  This difference 
is immaterial as far as we consider Young tableaux 
on a fixed Young diagram $\lambda$ to describe 
the associated Schur function $s_\lambda(\bsx)$.}
\beq
  T_{ij} > T_{i+1,j}, \quad 
  T_{ij} \ge T_{i,j+1}. 
\eeq
These semi-standard tableaux $T,T'$ encode 
the left and right halves of the plane partition. 
The Boltzmann weight $q^{|\pi|}$ thereby 
factorizes as 
\beq
  q^{|\pi|} = q^T q^{T'}, 
\eeq
where 
\beqnn
  q^T = \prod_{m=0}^\infty q^{(m+1/2)|\pi(-m)/\pi(-m-1)|},\quad 
  q^{T'} = \prod_{m=0}^\infty q^{(m+1/2)|\pi(m)/\pi(m+1)|}. 
\eeqnn

Since the triple $(\lambda,T,T')$ is in one-to-one 
correspondence with the plane partition $\pi$, 
the partition function $Z$ can be reorganized 
into the sum over the partition $\lambda = \pi(0)$ 
and the sum over the pair $(T,T')$ of 
semi-standard tableaux of shape $\lambda$: 
\beq
  Z = \sum_{\lambda}
      \sum_{T,T':\mathrm{shape}\,\lambda}q^Tq^{T'}. 
\eeq
By the well known combinatorial definition 
of the Schur functions, 
the partial sum over semi-standard tableaux 
becomes a special value of the Schur function 
$s_\lambda(\bsx)$ of infinite variables 
$\bsx = (x_1,x_2,\ldots)$ as 
\beq
    \sum_{T:\mathrm{shape}\,\lambda}q^T 
  = \sum_{T':\mathrm{shape}\,\lambda}q^{T'} 
  = s_\lambda(q^\rho), 
\eeq
where 
\beqnn
  q^\rho = (q^{1/2},q^{3/2},\ldots,q^{m+1/2},\ldots). 
\eeqnn
Note that the $q$-weights $q^T$ and $q^{T'}$ 
are identified with the monomials 
\beqnn
  \bsx^T = \prod_{(i,j)\in\lambda}x_{T_{ij}} 
\eeqnn
in the combinatorial definition 
\beqnn
  s_\lambda(\bsx) 
  = \sum_{T:\mathrm{shape}\,\lambda}\bsx^T 
\eeqnn
of Schur functions.  We thus obtain 
the Schur function expansion 
\beq
  Z = \sum_{\lambda}s_\lambda(q^\rho)^2 
\label{Z-Schur}
\eeq
of the partition function.

\subsection{Calculation of partition function}

We can calculate the sum (\ref{Z-Schur}) 
by the Cauchy identity 
\beq
  \sum_{\lambda}s_\lambda(\bsx)s_\lambda(\bsy) 
  = \prod_{i,j=1}^\infty (1 - x_iy_j)^{-1} 
  = \exp\left(\sum_{k=1}^\infty kt_k\bar{t}_k\right), 
\eeq
where 
\beqnn
  t_k = \frac{1}{k}\sum_{k=1}^\infty x_i^k, \quad 
  \bar{t}_k = \frac{1}{k}\sum_{k=1}^\infty y_i^k. 
\eeqnn
Letting $\bsx = \bsy = q^\rho$ amounts to setting 
\beqnn
  t_k = \bar{t}_k 
  = \frac{1}{k}\sum_{m=0}^\infty (q^{m+1/2})^k 
  = \frac{q^{k/2}}{k(1-q^k)}. 
\eeqnn
Consequently, 
\beqnn
\begin{aligned}
  \sum_\lambda s_\lambda(q^\rho)^2 
  &= \exp\left(\sum_{k=1}^\infty \frac{q^k}{k(1-q^k)^2}\right)\\
  &= \exp\left(\sum_{k=1}^\infty 
       \sum_{m,n=0}^\infty \frac{q^{mk+nk+k}}{k}\right) \\
  &= \exp\left(- \sum_{m,n=0}^\infty \log(1 - q^{m+n+1})\right)\\
  &= \prod_{m,n=0}^\infty (1 - q^{m+n+1})^{-1}. 
\end{aligned}
\eeqnn
Grouping the terms in the last infinite product 
with respect to the value of $l = m+n+1$, 
we find that the partition function becomes 
the so called MacMahon function: 
\beq
  Z = \prod_{l=1}^\infty (1 - q^l)^{-l}. 
\label{Z=MacMahon}
\eeq

\subsection{Models with external potentials}

The foregoing melting crystal model can be 
deformed by external potentials that depend 
on the main diagonal slice $\lambda = \pi(0)$.  

For example, we can insert the term $Q^{|\lambda|}$ 
($Q > 0$), namely, the potential $|\lambda|$ 
with coupling constant $\log Q$.  
The partition function 
\beqnn
  Z(Q) = \sum_\pi q^{|\pi|}Q^{|\pi(0)|} 
\eeqnn
can be calculated in much the same way 
and becomes the deformed MacMahon function 
\beq
  Z(Q) = \prod_{l=1}^\infty (1 - Qq^l)^{-l}. 
\eeq

We can further consider the potentials 
\beqnn
  \Phi_k(\lambda,p) = \sum_{i=1}^\infty q^{k(p+\lambda_i-i+1)} 
  - \sum_{i=1}^\infty q^{k(-i+1)}, 
\eeqnn
which depend on an integer parameter $p$ as well.  
These potentials originate in 5D supersymmetric 
$U(1)$ Yang-Mills theory \cite{MNTT04a,MNTT04b}.  
Note that this definition is rather heuristic; 
the infinite sums on the right hand side are divergent 
for the parameter $q$ in the range $0 < q < 1$.  
A true definition is obtained by by pairing the term 
$e^{p+\lambda_i-i+1}$ in the first sum 
with the term $e^{p-i+1}$ as 
\beq
  \Phi_k(\lambda,p) 
  = \sum_{i=1}^{\infty} (q^{k(p+\lambda_i-i+1)} - q^{k(p-i+1)}) 
    + q^k\frac{1-q^{pk}}{1-q^k}. 
\eeq
The second term on the right hand side 
amounts to the terms that cannot be paired. 
Since $\lambda_i = i$ for all but a finite number 
of $i$'s, the sum on the right hand side 
is a finite sum.  Thus $\Phi_i(\lambda,p)$ 
turns out to be well defined.  

Our previous work \cite{NT07} deals with the model 
deformed by the linear combination 
\beqnn
  \Phi(\bst,\lambda,p) 
  = \sum_{k=1}^\infty t_k\Phi_k(\lambda,p) 
\eeqnn
of these potentials with an infinite number 
of coupling constants $\bst = (t_1,t_2,\ldots)$.  
Its partition function reads 
\beqnn
  Z(\bst,p,Q) 
  = \sum_\pi q^{|\pi|}Q^{|\pi(0)|+p(p+1)/2}e^{\Phi(\bst,\pi(0),p)}. 
\eeqnn
Note that the potential $|\lambda|$ of 
the weight $Q^{|\lambda|}$ is also modified 
to $|\lambda| + p(p+1)/2$.  
Unlike $Z$ and $Z(Q)$, this partition function 
cannot be calculated in a closed form.  
We could, however, show that it coincides, 
up to a simple factor, with a tau function 
of the 1D Toda hierarchy.

\subsection{Models with two $q$-parameters}

We now turn to the models with two $q$-parameters 
\cite{IKS08}. These models are obtained by 
substituting  the Schur function factors as 
\beqnn
  s_\lambda(q^\rho)^2 \longrightarrow 
  s_\lambda(q_1^\rho)s_\lambda(q_2^\rho). 
\eeqnn
The new $q$-parameters $q_1,q_2$ are assumed 
to be in the range $0 < q_1 <1$ and $0 < q_2 < 1$.  
This substitution amounts to modifying 
the $q$-weights $q^T,q^{T'}$ of 
semi-standard tableaux $T,T'$ 
on the main diagonal slice as 
\beqnn
  q^T \longrightarrow q_1^T, \quad 
  q^{T'} \longrightarrow q_2^{T'}. 
\eeqnn

The previous partition functions $Z$, $Z(Q)$ 
and $Z(\bst,Q,p)$ are thereby replaced by 
\beqnn
\begin{aligned}
  Z(q_1,q_2) &= \sum_\lambda 
    s_\lambda(q_1^\rho)s_\lambda(q_2^\rho),\\
  Z(Q;q_1,q_2) &= \sum_\lambda 
    s_\lambda(q_1^\rho)s_\lambda(q_2^\rho)Q^{|\lambda|},\\
  Z(\bst,Q,p;q_1,q_2) &= \sum_\lambda 
    s_\lambda(q_1^\rho)s_\lambda(q_2^\rho)
    Q^{|\lambda|}e^{\Phi(\bst,\lambda,p)}. 
\end{aligned}
\eeqnn
Remember that the potentials $\Phi(\bst,\lambda,p)$ 
contain the third $q$-parameter $q$ as well.  
It is these partition functions that 
we shall consider in detail.  
As regards the first two, 
we can apply the previous method to 
give an infinite product formula: 
\beq
\begin{aligned}
  Z(q_1,q_2) &= \prod_{m,n=0}^\infty (1 - q_1^{m+1/2}q_2^{m+1/2})^{-1},\\
  Z(Q;q_1,q_2) &= \prod_{m,n=0}^\infty (1 - Qq_1^{m+1/2}q_2^{n+1/2})^{-1}. 
\end{aligned}
\eeq

\section{Fermionic formula of partition function}

We use a 2D complex free fermion system 
to reformulate the partition functions.  
Notations and conventions are the same 
as those in our previous work \cite{NT07}.

\subsection{Complex fermions}

Let $\psi_n,\psi^*_n$ ($n \in \ZZ$) denote 
the Fourier modes of the free fermion fields 
\beqnn
  \psi(z) = \sum_{n=-\infty}^\infty \psi_nz^{-n-1}, \quad 
  \psi^*(z) = \sum_{n=-\infty}^\infty \psi^*_nz^{-n}. 
\eeqnn
They satisfy the anti-commutation relations 
\beqnn
  \{\psi_m,\psi^*_n\} = \delta_{m+n,0}, \quad 
  \{\psi_m,\psi_n\} = \{\psi^*_m,\psi^*_n\} = 0. 
\eeqnn
The Fock space $\mathcal{F}$ splits into 
charge $p$ subspaces $\mathcal{F}_p$ ($p \in \ZZ$).  
$\mathcal{F}_p$ has a normalized ground state 
(charge $p$ vacuum) $|p\rangle$, which is characterized 
by the vacuum condition 
\beqnn
  \psi_n|p\rangle = 0 \quad\mbox{for}\quad n \ge -p, 
  \qquad 
  \psi^*_n|p\rangle = 0 \quad\mbox{for}\quad n \ge p+1. 
\eeqnn 
The dual Fock space $\mathcal{F}^*$, too, splits into 
charge $p$ subspaces $\mathcal{F}^*_p$ ($p \in \ZZ$). 
The vacuum condition for the normalized ground state 
$\langle p|$ of $\mathcal{F}^*_p$ reads 
\beqnn
  \langle p|\psi_n = 0 \quad\mbox{for}\quad n \le -p-1, 
  \qquad 
  \langle p|\psi^* = 0 \quad\mbox{for}\quad n \le p.
\eeqnn
$\mathcal{F}_p$ and $\mathcal{F}^*_p$ have excited states 
$|\lambda,p\rangle$ and $\langle\lambda,p|$ 
labeled by partition $\lambda$, which altogether 
form a basis of $\mathcal{F}_p$ and $\mathcal{F}^*_p$.  
If $\lambda$ is of length $\le n$, namely, 
$\lambda = (\lambda_1,\lambda_2,\ldots,\lambda_n,0,0,\ldots)$, 
these excited states can be obtained from $|p\rangle$ as 
\beqnn
\begin{aligned}
  |\lambda,p\rangle 
  &= \psi_{-(p+\lambda_1-1)-1}\cdots\psi_{-(p+\lambda_n-n)-1}
     \psi^*_{(p-n)+1}\cdots\psi^*_{(p-1)+1}|p\rangle, \\
  \langle\lambda,p| 
  &= \langle p|\psi_{-(p-1)-1}\cdots\psi_{-(p-n)-1}
     \psi^*_{(p+\lambda_n-n)+1}\cdots\psi^*_{(p+\lambda_1-1)+1}. 
\end{aligned}
\eeqnn
These states are mutually orthonormal: 
\beqnn
  \langle\lambda,p|\mu,q\rangle 
  = \delta_{\lambda\mu}\delta_{pq}. 
\eeqnn

\subsection{Partition functions of simplest model}

Let us introduce the special fermion bilinears 
\beqnn
\begin{aligned}
  J_m = \sum_{n=-\infty}^\infty {:}\psi_{m-n}\psi^*_n{:},\quad 
  L_0 = \sum_{n=-\infty}^\infty n{:}\psi_{-n}\psi^*_n{:},\quad 
  H_k = \sum_{n=-\infty}^\infty q^{kn}{:}\psi_{-n}\psi^*_n{:},
\end{aligned}
\eeqnn
where ${:}\quad{:}$ stands for normal ordering, namely, 
\beqnn
  {:}\psi_m\psi^*_n{:} 
  = \psi_m\psi^*_n - \langle 0|\psi_m\psi^*_n|0\rangle. 
\eeqnn
$J_m$'s are the Fourier modes of the $U(1)$ current 
\beqnn
  J(z) = {:}\psi(z)\psi^*(z){:} 
  = \sum_{m=-\infty}^\infty J_mz^{-m-1}, 
\eeqnn
and satisfy the commutation relations 
\beq
  [J_m,J_n] = m\delta_{mn}
\eeq
of the Heisenberg algebra.  $L_0$ is one of 
the basis $\{L_m\}_{m=-\infty}^\infty$ of 
the Virasoro algebra.  $H_k$'s are the same 
``Hamiltonians'' as used in the case of 
$q_1 = q_2 = 1$ \cite{NT07}.  
$H_k$'s and $L_0$ commutate with each other, 
and have the excited states $\langle\lambda,p|$ 
and $|\lambda,p\rangle$ as joint eigenstates, 
namely, 
\beq
  \langle\lambda,p|H_k 
= \langle\lambda,p|\Phi_k(\lambda,p)|, \quad 
  \langle\lambda,p|L_0 
= \langle\lambda,p|\left(|\lambda| + \frac{p(p+1)}{2}\right) 
\eeq
and 
\beq
  H_k|\lambda,p\rangle 
= \Phi_k(\lambda,p)|\lambda,p\rangle, \quad 
  L_0|\lambda,p\rangle 
= \left(|\lambda| + \frac{p(p+1)}{2}\right)|\lambda,p\rangle.
\eeq
In particular, the potentials $\Phi_k(\lambda,p)$ 
and their linear combination $\Phi(\bst,\lambda,p)$ 
show up here as the eigenvalues of $H_k$ and 
their linear combinations 
\beqnn
  H(\bst) = \sum_{k=1}^\infty t_kH_k. 
\eeqnn

We further introduce the exponential operators \cite{ORV03}
\beqnn
  G_{\pm} = \exp\left(\sum_{k=1}^\infty 
              \frac{q^{k/2}}{k(1-q^k)}J_{\pm k}\right), 
\eeqnn
which belong to the Clifford group $GL(\infty)$. 
These operators can be factorized as 
\beq
  G_{\pm} = \prod_{m=0}^\infty V_{\pm}(q^{m+1/2}), 
\eeq
where $V_{\pm}(z)$ denote the familiar vertex operators 
\beqnn
  V_{\pm}(z) 
  = \exp\left(\sum_{k=1}^\infty \frac{z^k}{k}J_{\pm k}\right). 
\eeqnn
These vertex operators act on $\langle \lambda,p|$ 
and $|\lambda,p\rangle$ as 
\beq
  \langle\lambda,p| V_{+}(z) 
= \sum_{\mu\succ\lambda}z^{|\mu|-|\lambda|}\langle\mu,p|,\quad
  V_{-}(z)|\lambda,p\rangle  
= \sum_{\mu\succ\lambda}z^{|\mu|-|\lambda|}|\mu,p\rangle. 
\eeq
One can thereby deduce \cite{ORV03} that 
the action of $G_{\pm}$ on the ground states 
$\langle p|$ and $|p\rangle$ yields 
a linear combination of $\langle\lambda,p|$ 
and $|\lambda,p\rangle$ with coefficients 
$s_\lambda(q^\rho)$: 
\beq
\begin{aligned}
& \langle p|G_{+} 
  = \sum_{\lambda}\sum_{T:\mathrm{shape}\,\lambda}q^T 
    \langle \lambda,p| 
  = \sum_{\lambda} s_\lambda(q^\rho) \langle \lambda,p|, \\
& G_{-}|p\rangle 
  = \sum_{\lambda}\sum_{T':\mathrm{shape}\,\lambda}q^{T'} 
    |\lambda,p\rangle 
  = \sum_{\lambda} s_\lambda(q^\rho) |\lambda,p\rangle.
\end{aligned}
\label{PP-state}
\eeq
$G_{\pm}$ thus play the role of ``transfer 
(or transition) matrices'' in lattice models.  

Since $\langle\lambda,p|$ and $|\lambda,p\rangle$ 
are orthonormal, the inner product of 
$\langle p|G_{+}$ and $G_{-}|p\rangle$ 
becomes a sum of $s_\lambda(q^\rho)^2$, 
\beqnn
  \langle p| G_{+}G_{-} |p\rangle 
  = \sum_{\lambda,\mu}
    s_\lambda(q^\rho)s_\mu(q^\rho)\delta_{\lambda\mu}
  = \sum_\lambda s_\lambda(q^\rho)^2, 
\eeqnn
which is independent of $p$.  
We thus obtain the fermionic formula 
\beq
  Z = \langle 0|G_{+}G_{-}|0\rangle
\label{Z-fermion}
\eeq
of the partition function $Z$.  
The MacMahon function formula (\ref{Z=MacMahon}) 
can be also derived from from this formula 
and the commutation relation 
\beq
  G_{+}G_{-} 
  = G_{-}G_{+}\exp\left(\sum_{k=1}^\infty\frac{q^k}{k(1-q^k)^2}\right) 
\eeq
of $G_{\pm}$.

\subsection{Partition functions of deformed models}

A fermionic formula of $Z(Q)$ and $Z(\bst,Q,p)$ 
can be obtained by inserting a combination of 
$L_0$ and $H_k$'s with coupling constants 
into  (\ref{Z-fermion}).  As regards $Z(Q)$, 
the operator to be inserted is $Q^{L_0}$, 
which acts on $\langle\lambda,0|$ 
and $|\lambda,0\rangle$ as 
\beqnn
  \langle\lambda,0|Q^{L_0} = \langle\lambda,0|Q^{|\lambda|},\quad
  Q^{L_0}|\lambda,0\rangle = Q^{|\lambda|}|\lambda,0\rangle. 
\eeqnn
$\langle 0|G_{+}Q^{L_0}G_{-}|0 \rangle$ 
can be thereby expanded as 
\beqnn
  \langle 0|G_{+}Q^{L_0}G_{-}|0 \rangle
  = \sum_{\lambda} 
    \langle 0|G_{+}Q^{L_0}|\lambda,0\rangle
    \langle\lambda,0|G_{-}|0\rangle 
  = \sum_\lambda s_\lambda(q^\rho)^2Q^{|\lambda|}. 
\eeqnn
This is nothing but $Z(Q)$. Thus we obtain 
the fermionic formula 
\beq
  Z(Q) = \langle 0|G_{+}Q^{L_0}G_{-}|0 \rangle.
\eeq
In the same way, we can derive the fermionic formula 
\beq
  Z(\bst,Q,p) = \langle p| G_{+}Q^{L_0}e^{H(\bst)}G_{-} |p\rangle 
\eeq
for $Z(\bst,Q,p)$.  

These fermionic formulae can be readily 
generalized to $Z(q_1,q_2)$, $Z(Q;q_1,q_2)$ 
and $Z(\bst,Q,p;q_1,q_2)$. 
Let us introduce the operators 
\beqnn
\begin{aligned}
  G_{+}(q_1) &= \exp\left(\sum_{k=1}^\infty 
                  \frac{q_1^{k/2}}{k(1-q_1^k)}J_k\right), \\
  G_{-}(q_2) &= \exp\left(\sum_{k=1}^\infty 
                  \frac{q_2^{k/2}}{k(1-q_2^k)}J_{-k}\right) 
\end{aligned}
\eeqnn
in place of $G_{\pm}$.  As shown in (\ref{PP-state}), 
these operators generate a linear combination 
of $\langle \lambda,p|$ and $|\lambda,p\rangle$ 
with coefficients $s_\lambda(q_1^\rho)$ 
and $s_\lambda(q_2^\rho)$.  
The rest of calculations in the previous case 
applies to this case without modification.  
Thus we obtain the following fermionic formulae 
of $Z(q_1,q_2)$, $Z(Q;q_1,q_2)$ and $Z(\bst,Q,p;q_1,q_2)$: 
\beq
\begin{aligned}
  Z(q_1,q_2) &= \langle 0|G_{+}(q_1)G_{-}(q_2)|0\rangle, \\
  Z(Q;q_1,q_2) &= \langle 0|G_{+}(q_1)Q^{L_0}G_{-}(q_2)|0\rangle,\\
  Z(\bst,Q,p;q_1,q_2) 
    &= \langle p|G_{+}(q_1)Q^{L_0}e^{H(\bst)}G_{-}(q_2)|p\rangle. 
\end{aligned}
\label{Z(q1,q2)-fermion}
\eeq

\section{Relation to Toda hierarchy}

We now turn to the issues of integrable structures. 
It is shown in our previous paper \cite{NT07} 
that the partition function $Z(\bst,Q,p)$ 
coincides, up to a simple factor, with 
a special tau function of the 1D Toda hierarchy. 
The goal of this section is to generalize 
this result to the case where $q_1$ and $q_2$ 
are related to $q$ as 
\beq
  q_1 = q^{1/N_1}, \quad 
  q_2 = q^{1/N_2} 
  \label{q1-q2-q}
\eeq
for some positive integers $N_1$ and $N_2$.  
We assume this condition throughout this section.

\subsection{Intertwining relations in quantum torus algebra}

Let us recall yet another set of fermion bilinears 
\beqnn
  W_0 = \sum_{n=-\infty}^\infty n^2{:}\psi_{-n}\psi^*_n{:},\quad 
  V^{(k)}_m 
  = q^{-km/2}\sum_{n=-\infty}^\infty q^{kn}{:}\psi_{m-n}\psi^*_n{:} 
\eeqnn
from our previous work \cite{NT07}.  
$W_0$ is one of the basis $\{W_m\}_{m=-\infty}^\infty$ 
of the so called $W_3$ subalgebra 
in the $W_\infty$ algebra of complex fermions.  
$V^{(k)}_m$'s satisfy the commutation relations 
\beq
  [V^{(k)}_m,\, V^{(l)}_n] 
  = (q^{(lm-kn)/2}-q^{(kn-lm)/2})
    (V^{(k+l)}_{m+n} - \delta_{m+n,0}\frac{q^{k+l}}{1-q^{k+l}}) 
\eeq
of the quantum torus algebra.%
\footnote{Substantially the same realization of 
this algebra is considered in different contexts 
by Gao \cite{Gao02} and Okounkov and Pandharipande 
\cite{OP02a,OP02b}.}   Actually, we need 
just a {\it half} of this algebra (so to speak, 
a {\it quantum cylinder algebra}) 
spanned by $V^{(k)}_m$ with $k \ge 1$ and $m \in \ZZ$.  
$J_m$ and $H_k$ are part of this basis: 
\beq
  J_m = V^{(0)}_m, \quad 
  H_k = V^{(k)}_0. 
\eeq
This Lie algebra has an inner symmetry 
(``shift symmetry'') \cite{NT07}.  
A consequence of this symmetry is the following 
intertwining relations among $J_m$'s and $H_k$'s, 
which play a fundamental role in identifying 
the integrable structure  of $Z(\bst,Q,p)$. 

\begin{lemma}
$J_m$'s and $H_k$'s satisfy the intertwining relations 
\beq
\begin{aligned}
  q^{W_0/2}G_{-}G_{+}H_k 
  &= \left((-1)^kJ_k + \frac{q^k}{1-q^k}\right)q^{W_0/2}G_{-}G_{+}, \\
  H_kG_{-}G_{+}q^{W_0/2} 
  &= G_{-}G_{+}q^{W_0/2}\left((-1)^lJ_{-k} + \frac{q^k}{1-q^k}\right). 
\end{aligned}
\label{intertwining}
\eeq
\end{lemma}

We now use these intertwining relations 
replacing $q \to q_1,q_2$.  (\ref{q1-q2-q}) 
implies that $H_k$ may be thought of 
as elements of the quantum torus algebras 
with $q$-parameters $q = q_1$ and $q = q_2$, 
namely, 
\beq
  H_k = V^{(N_1k)}_0(q_1) = V^{(N_2k)}_0(q_2), 
\eeq
where $V^{(k)}_m(q_1)$ and $V^{(k)}_m(q_2)$ 
denote the counterparts of $V^{(k)}_m$ 
for $q = q_1$ and $q = q_2$.  Therefore, 
applying (\ref{intertwining}) to the cases 
where $q = q_1$ and $q = q_2$, 
we obtain the relations 
\beq
  q_1^{W_0/2}G_{-}(q_1)G_{+}(q_1)H_k 
  = \left((-1)^{N_1k}J_{N_1k} + \frac{q^k}{1-q^k}\right) 
    q_1^{W_0/2}G_{-}(q_1)G_{+}(q_1) 
\label{intertwining1}
\eeq
interchanging $J_{N_1k}$'s and $H_k$'s and 
\beq
  H_kG_{-}(q_2)G_{+}(q_2)q_2^{W_0/2} 
  = G_{-}(q_2)G_{+}(q_2)q_2^{W_0/2}
    \left((-1)^{N_2k}J_{-N_2k} + \frac{q^k}{1-q^k}\right) 
\label{intertwining2}
\eeq
interchanging $J_{-N_2k}$'s and $H_k$'s.  
Note that the c-number terms in the parentheses 
have been rewritten as 
\beq
    \frac{q_1^{N_1k}}{1-q^{N_1k}} 
  = \frac{q_2^{N_2k}}{1-q^{N_2k}} 
  = \frac{q^k}{1-q^k}. 
\eeq

\subsection{Partition function as tau function}

Armed with these intertwining relations, 
we can generalize our previous result \cite{NT07} 
to $Z(\bst,Q,p;q_1,q_2)$.  
Let us recall that a general tau function 
$\tau(\bsT,\bsTbar,p)$ of the Toda hierarchy 
depends on two sets of continuous variables 
$\bsT = (T_1,T_2,\ldots)$ and 
$\bsTbar = (\Tbar_1,\Tbar_2,\ldots)$ 
and has a fermionic formula of the form 
\cite{JM83,Takebe91} 
\beq
  \tau(\bsT,\bsTbar,p) 
  = \langle p|\exp\left(\sum_{k=1}^\infty T_kJ_k\right) 
    g \exp\left(- \sum_{k=1}^\infty \Tbar_kJ_{-k}\right) 
    |p \rangle, 
\label{tau(T,Tbar,p)-fermion}
\eeq
where $g$ is an element of $GL(\infty)$. 
We now show that $Z(\bst,Q,p;q_1,q_2)$ coincides, 
up to a simple factor, with such a tau function. 

\begin{theorem}
$Z(\bst,Q,p;q_1,q_2)$ can be rewritten 
in two different forms as 
\begin{multline}
Z(\bst,Q,p;q_1,q_2) 
  = (q_1q_2)^{-p(p+1)(2p+1)/12} 
     \exp\left(\sum_{k=1}^\infty \frac{q^kt_k}{1-q^k}\right) \\
  \mbox{}\times 
  \langle p|\exp\left(\sum_{k=1}^\infty 
    (-1)^{N_1k}t_kJ_{N_1k}\right) g |p\rangle
\label{Z(q1,q2)=tau1}
\end{multline}
and 
\begin{multline}
Z(\bst,Q,p;q_1,q_2) 
  =  (q_1q_2)^{-p(p+1)(2p+1)/12} 
     \exp\left(\sum_{k=1}^\infty \frac{q^kt_k}{1-q^k}\right) \\
  \mbox{}\times 
  \langle p| g\exp\left(\sum_{k=1}^\infty 
    (-1)^{N_2k}t_kJ_{-N_2k}\right) |p\rangle, 
\label{Z(q1,q2)=tau2}
\end{multline}
where $g$ is an element of $GL(\infty)$ of the form 
\beq
  g = q_1^{W_0/2}G_{-}(q_1)G_{+}(q_1)Q^{L_0} 
      G_{-}(q_2)G_{+}(q_2)q_2^{W_0/2}. 
\label{g-for-Z(q1,q2)}
\eeq
\end{theorem}

\proof 
To derive (\ref{Z(q1,q2)=tau1}),  let us think of 
the right hand side of the fermionic formula 
(\ref{Z(q1,q2)-fermion}) as the inner product of 
$\langle p|G_{+}(q_1) \in \mathcal{F}^*$ 
and $G_{-}(q_2)|p\rangle \in \mathcal{F}$ 
in which $Q^{L_0}e^{H(\bst)}$ is inserted.  
Also remember that the order of 
$Q^{L_0}$ and $e^{H(\bst)}$ is immaterial, 
because $L_0$ and $H(\bst)$ commute with each other.  
Since $G_{-}(q_1)$ and $q_1^{W_0/2}$ act 
on $\langle p|$ almost trivially as 
\beqnn
  \langle p|G_{-}(q_1) = \langle p|, \quad 
  \langle p|q^{W_0/2} = q_1^{p(p+1)(2p+1)/12}\langle p|, 
\eeqnn
we can rewrite $\langle p|G_{+}(q_1)$ 
in (\ref{Z(q1,q2)-fermion}) as 
\beqnn
  \langle p|G_{+}(q_1)
  = q_1^{-p(p+1)(2p+1)/12}
    \langle p|q_1^{W_0/2}G_{-}(q_1)G_{+}(q_1). 
\eeqnn
In the same way, 
\beqnn
  G_{-}(q_2)|p\rangle 
  = q_2^{-p(p+1)(2p+1)/12} 
    G_{-}(q_2)G_{+}(w_2)q_2^{W_0/2}|p\rangle. 
\eeqnn
Thus $Z(\bst,Q,p;q_1,q_2)$ can be cast into 
such a form as 
\begin{multline*}
  Z(\bst,Q,p;q_1,q_2) 
  = (q_1q_2)^{-p(p+1)(2p+1)/12} \\
  \mbox{}\times 
  \langle p|q_1^{W_0/2}G_{-}(q_1)G_{+}(q_1)e^{H(\bst)} 
    Q^{L_0}G_{-}(q_2)G_{+}(q_2)q_2^{W_0/2}|p\rangle. 
\end{multline*}
Since the intertwining relation (\ref{intertwining1}) 
implies the identity 
\begin{multline*}
  q_1^{W_0/2}G_{-}(q_1)G_{+}(q_1)e^{H(\bst)} \\
= \exp\left(\sum_{k=1}^\infty\frac{q^kt_k}{1-q^k}\right) 
  \exp\left(\sum_{k=1}^\infty (-1)^{N_1k}t_kJ_{N_1k}\right) 
  q_1^{W_0/2}G_{-}(q_1)G_{+}(q_1), 
\end{multline*}
we can move $e^{H(\bst)}$ to the right of $\langle p|$ 
and obtain the first formula (\ref{Z(q1,q2)=tau1}).  
The second formula (\ref{Z(q1,q2)=tau2}) 
can be derived in much the same way. 
\qed
\bigskip

Thus $Z(\bst,Q,p;q_1,q_2)$ turns out to coincide, 
up to a simple factor, with the tau function 
$\tau(\bsT,\bsTbar,p)$ determined by 
the $GL(\infty)$ element $g$ of (\ref{g-for-Z(q1,q2)}).  
Note that the time variables $\bsT,\bsTbar$ 
of the full Toda hierarchy are now restricted 
to a subspace.  In particular, unlike the case 
where $N_1 = N_2 = 1$, not all of these variables 
join the game.  Namely, it is only $T_{N_1k}$ 
and $\Tbar_{N_2k}$, $k = 1,2,\ldots$, 
that correspond to the coupling constants $\bst$ 
of this generalized melting crystal model.

\subsection{Bigraded Toda hierarchy}

Since the same partition function 
$Z(\bst,Q,p;q_1,q_2)$ is expressed in 
two apparently different forms as 
(\ref{Z(q1,q2)=tau1}) and (\ref{Z(q1,q2)=tau2}), 
we find that the identity 
\beq
  \langle p|\exp\left(\sum_{k=1}^\infty 
   (-1)^{N_1k}t_kJ_{N_1k}\right) g |p\rangle
=  \langle p| g\exp\left(\sum_{k=1}^\infty 
   (-1)^{N_2k}t_kJ_{-N_2k}\right) |p\rangle 
\eeq
holds.  This is a manifestation of the following 
more fundamental fact.  

\begin{theorem}
$J_k$'s and the $GL(\infty)$ element $g$ 
of (\ref{g-for-Z(q1,q2)}) satisfy 
the intertwining relations 
\beq
  (-1)^{N_1k}J_{N_1k}g = g(-1)^{N_2k}J_{-N_2k} 
\label{Jg=gJ}
\eeq
for $k = 1,2,\ldots$.  
\end{theorem}

\proof
Using the intertwining relations (\ref{intertwining1}) 
and (\ref{intertwining2}), we can derive (\ref{Jg=gJ}) 
as follows: 
\beqnn
\begin{aligned}
(-1)^{N_1k}J_{N_1k}g 
  &= (-1)^{N_1k}J_{N_1k}q_1^{W_0/2}G_{-}(q_1)G_{+}(q_1) 
     Q^{L_0}G_{-}(q_2)G_{+}(q_2)q_2^{W_0/2} \\
  &= q_1^{W_0/2}G_{-}(q_1)G_{+}(q_1) 
     \left(H_k - \frac{q^k}{1-q^k}\right) 
     Q^{L_0}G_{-}(q_2)G_{+}(q_2)q_2^{W_0/2} \\
  &= q_1^{W_0/2}G_{-}(q_1)G_{+}(q_1)Q^{L_0} 
     \left(H_k - \frac{q^k}{1-q^k}\right) 
     G_{-}(q_2)G_{+}(q_2)q_2^{W_0/2} \\
  &= q_1^{W_0/2}G_{-}(q_1)G_{+}(q_1)Q^{L_0} 
     G_{-}(q_2)G_{+}(q_2)q_2^{W_0/2}
     (-1)^{N_2k}J_{-N_2k} \\
  &= g(-1)^{N_2k}J_{-N_2k}. 
\end{aligned}
\eeqnn
\qed
\bigskip

By these intertwining relations, 
we can freely move $J_{N_1k}$ and $J_{-N_2k}$ 
to the far side of $g$.  This implies 
that $\tau(\bsT,\bsTbar,p)$ depends on 
$T_{N_1k}$ and $\Tbar_{N_2k}$ only through 
the linear combination 
$(-1)^{N_1k}T_{N_1k} - (-1)^{N_2k}\Tbar_{N_2k}$.
In other words, $\tau(\bsT,\bsTbar,p)$ 
satisfies the constraints 
\beq
  (-1)^{N_1k}\frac{\rd\tau}{\rd T_{N_1k}} 
  + (-1)^{N_2k}\frac{\rd\tau}{\rd \Tbar_{N_2k}} = 0 
\label{tau-constraint}
\eeq
for $k = 1,2,\ldots$.  
Apart from the presence of the signature factors 
$(-1)^{N_1k}$ and $(-1)^{N_2k}$, 
these constraints are the same as those 
that characterize the bigraded Toda hierarchy 
of type $(N_1,N_2)$ \cite{Carlet06} 
as a reduction of the full Toda hierarchy.  

In the language of the Lax operators \cite{UT84,TT95}
\beqnn
  L = e^{\rd_p} + u_1 + u_2e^{-\rd_p} + \cdots, \quad 
  \Lbar = \ubar_0e^{\rd_p} + \ubar_1e^{2\rd_p} + \cdots 
\eeqnn
of the Toda hierarchy, the reduction to 
the (slightly modified) bigraded Toda hierarchy 
can be characterized by the constraint 
\beq
  (-L)^{N_1} = (-\Lbar)^{-N_2}.  
\label{L-constraint}
\eeq
Let $\mathcal{L}$ denote the difference operator 
defined by both hand sides of this constraint.  
This reduced Lax operator is a difference operator 
of the form 
\beq
  \mathcal{L} 
  = (-e^{\rd_p})^{N_1} + b_1(-e^{\rd_p})^{N_1-1} 
    + \cdots + b_{N_1+N_2}(-e^{\rd_p})^{-N_2}, 
\eeq
and satisfies the Lax equations 
\beq
  \frac{\rd\mathcal{L}}{\rd T_k} = [B_k,\mathcal{L}], \quad 
  \frac{\rd\mathcal{L}}{\rd \Tbar_k} = [\Bbar_k,\mathcal{L}], 
\eeq
where $B_k$ and $\Bbar_k$ are given by 
\beqnn
  B_k = (L^k)_{\ge 0}, \quad 
  \Bbar_k = (\Lbar^{-k})_{<0}, 
\eeqnn
$(\quad)_{\ge 0}$ and $(\quad)_{<0}$ standing 
for the projection onto the part of 
nonnegative and negative powers of $e^{\rd_p}$. 
More precisely, since the $(\quad)_{\ge 0}$ and 
the $(\quad)_{<0}$ parts of $\mathcal{L} 
= (-L)^{N_1} = (-\Lbar)^{-N_2}$ are given by 
\beqnn
  (\mathcal{L})_{\ge 0} = (-1)^{N_1}B_{N_1}, \quad 
  (\mathcal{L})_{<0} = (-1)^{N_2}\Bbar_{N_2}, 
\eeqnn
we can express $\mathcal{L}$ itself as 
\beq
  \mathcal{L} = (-1)^{N_1}B_{N_1} + (-1)^{N_2}\Bbar_{N_2}. 
\eeq
The powers $\mathcal{L}^k$, $k = 1,2,\ldots$, 
of $\mathcal{L}$ can be likewise expressed as 
\beqnn
  \mathcal{L}^k = (-1)_{N_1k}B_{N_1k} + (-1)^{N_2k}\Bbar_{N_2k}. 
\eeqnn
This readily implies that the stationary equations 
\beq
  (-1)^{N_1k}\frac{\rd\mathcal{L}}{\rd T_{N_1k}} 
  + (-1)^{N_2k}\frac{\rd\mathcal{L}}{\rd \Tbar_{N_2k}} 
  = [\mathcal{L}^k,\mathcal{L}] 
  = 0, 
\eeq
hold for $\mathcal{L}$.  These stationary equations 
are counterparts of (\ref{tau-constraint}) 
in the Lax formalism.

\section{Relation to $q$-difference Toda equation}

In this section, we consider another integrable structure, 
which is hidden in $Q$-dependence of the partition function. 
We no longer have to assume (\ref{q1-q2-q}), namely, 
$q_1,q_2$ and $q$ are independent parameters 
throughout this section.  

\subsection{Variables for $q$-difference analogue}

It is known \cite{MMV94,AHvM98,Takasaki05} 
that tau functions of the Toda hierarchy can be 
converted to tau functions of a $q$-difference 
analogue of the 2D Toda equation 
by changing variables as 
\footnote{This definition is slightly modified 
from the common one in the literature 
\cite{MMV94,AHvM98,Takasaki05}.  
We shall make a remark on this issue later on.}
\beq
  T_k = \frac{x^k}{k(1-q_1^k)}, \quad 
  \Tbar_k = - \frac{y^k}{k(1-q_2^k)}. 
\label{TTbar-xy}
\eeq
Let $\sigma(x,y,p)$ denote such a transformed tau function, 
namely, 
\beq
  \sigma(x,y,p) = \tau([x]_{q_1}, -[y]_{q_2},p), 
\label{sigma(x,y,p)-tau}
\eeq
where $\tau(\bsT,\bsTbar,p)$ is a tau function 
of the Toda hierarchy and $[x]_q$ denotes 
the $q$-difference analogue 
\beqnn
  [x]_q = \left(\frac{x}{1-q},\frac{x^2}{2(1-q^2)},\ldots, 
          \frac{x^k}{k(1-q^k)},\ldots\right) 
\eeqnn
of the notation 
\beqnn
  [x] = \left(x,\frac{x^2}{2},\ldots,\frac{x^k}{k},\ldots\right) 
\eeqnn
that play a fundamental role in the study of 
KP and Toda hierarchies.  

By this change of variables (\ref{TTbar-xy}), 
the fermionic formula (\ref{tau(T,Tbar,p)-fermion}) 
of tau functions of the Toda hierarchy 
turns into the formula 
\beq
  \sigma(x,y,p) 
  = \langle p|
    \exp\left(\sum_{k=1}^\infty\frac{x^k}{k(1-q_1^k)}J_k\right)g
    \exp\left(\sum_{k=1}^\infty\frac{y^k}{k(1-q_2^k)}J_{-k}\right)
    |p \rangle 
\label{sigma(x,y,p)-fermion}
\eeq
of $\sigma(x,y,p)$.  In particular, when $x,y$ and $g$ 
are specialized as 
\beq
  x = q_1^{k/2}, \quad y = q_2^{k/2}, \quad 
  g = Q^{L_0}e^{H(\bst)}, 
\label{special-xyg}
\eeq
$\sigma(x,y,p)$ coincides with $Z(\bst,Q,p;q_1,q_2)$. 
The goal of this section is to derive a $q$-difference 
equation that $Z(\bst,Q,p;q_1,q_2)$ satisfies 
with respect to $Q$.

\subsection{$q$-difference 2D Toda equation for $\sigma(x,y,p)$}

As a preliminary step towards a $q$-difference equation 
for $Z(\bst,Q,p;q_1,q_2)$, we now show that 
the transformed tau function $\sigma(x,y,t)$ 
satisfies a $q$-difference analogue of 
the 2D Toda equation (in a bilinear form) 
\begin{multline}
  \frac{\rd^2\tau(x,y,p)}{\rd x\rd y}\tau(x,y,p) 
  - \frac{\rd\tau(x,y,p)}{\rd x}\frac{\rd\tau(x,y,p)}{\rd y}\\
  = \tau(x,y,p+1)\tau(x,y,p-1). 
\label{2DToda}
\end{multline}
Now that this is the lowest equation satisfied 
by the tau function of the Toda hierarchy 
with respect to $x = T_1$ and $y = - \Tbar_1$.  

\begin{lemma}
For any tau function of the Toda hierarchy, 
the function $\sigma(x,y,p)$ defined by 
(\ref{sigma(x,y,p)-tau}) satisfies 
the $q$-difference 2D Toda equation 
\begin{multline}
  \sigma(q_1x,q_2y,p)\sigma(x,y,p) - \sigma(x,q_2y,p)\sigma(q_1x,y,p) \\
  = xy\sigma(x,y,p+1)\sigma(q_1x,q_2y,p-1). 
\label{q-diff-2DToda}
\end{multline}
\end{lemma}

\proof
This is a consequence of the difference analogue 
\begin{multline}
  \tau(\bsT-[x],\bsTbar,p)\tau(\bsT,\bsTbar-[y],p) 
  - \tau(\bsT,\bsTbar,p)\tau(\bsT-[x],\bsTbar-[y],p) \\
  = xy\tau(\bsT,\bsTbar-[y],p+1)\tau(\bsT-[x],\bsT,p-1) 
\label{diff-Fay}
\end{multline}
of the 2D Toda equation (\ref{2DToda}).  This equation 
is one of ``Fay-type identities'' \cite{Teo06,Takasaki07} 
that hold for any tau function of the Toda hierarchy.  
We shift $\bsTbar$ as $\bsTbar \to \bsTbar + [y]$ 
and substitute  $\bsT = [x]_{q_1}$ and 
$\bsTbar = - [y]_{q_2}$ in this equation.  
The outcome is the equation 
\begin{multline*}
  \tau([x]_{q_1}-[x],-[y]_{q_2}+[y],p)\tau([x]_{q_1},-[y]_{q_2},p)\\
  - \tau([x]_{q_1},-[y]_{q_2}+[y],p)\tau([x]_{q_1}-[x],-[y]_{q_2},p)\\
  = xy\tau([x]_{q_1},-[y]_{q_2},p+1)\tau([x]_{q_1}-[x],-[y]_{q_2}+[y],p-1), 
\end{multline*}
which we can further rewrite as 
\begin{multline*}
  \tau([q_1x]_{q_1},-[q_2y]_{q_2},p)\tau([x]_{q_1},-[y]_{q_2},p)\\
  - \tau([x]_{q_1},-[q_2y]_{q_2},p)\tau([q_1x]_{q_1},-[y]_{q_2},p)\\
  = xy\tau([x]_{q_1},-[y]_{q_2},p+1)\tau([q_1x]_{q_1},-[q_2y]_{q_2},p-1) 
\end{multline*}
by the $q$-shift property 
\beq
  [qx]_{q} = [x]_{q} - [x] 
\eeq
of $[x]_q$.  The last equation is nothing but 
(\ref{q-diff-2DToda}).  
\qed
\bigskip

A few remarks are in order.  

\begin{itemize}
\item[1.] One can rewrite the $q$-difference 
equation (\ref{q-diff-2DToda}) as 
\begin{multline}
  D_{q_1,x}D_{q_2,y}\sigma(x,y,p)\cdot \sigma(x,y,p) 
  - D_{q_1,x}\sigma(x,y,p)\cdot D_{q_2,y}\sigma(x,y,p) \\
  = \sigma(x,y,p+1)\sigma(q_1,q_2y,p-1), 
\end{multline}
where $D_{q_1,x}$ and $D_{q_2,y}$ stand for 
the $q$-difference operators 
\beqnn
\begin{aligned}
  D_{q_1,x}\sigma(x,y,p) 
&= \frac{\sigma(x,y,p)-\sigma(q_1x,y,p)}{(1-q_1)x}, \\
  D_{q_2,y}\sigma(x,y,p) 
&= \frac{\sigma(x,y,p)-\sigma(x,q_2y,p)}{(1-q_2)y}. 
\end{aligned}
\eeqnn
As $q_1,q_2 \to 0$, this equation 
turns into the 2D Toda equation (\ref{2DToda}). 
\item[2.]
One can modify (\ref{TTbar-xy}) as 
\beq
  T_k = \frac{x^k}{k(1-q_1^k)}, \quad 
  \Tbar_k = \frac{y^k}{k(1-q_2^k)}. 
\eeq
The transformed tau function 
\beq
  \rho(x,y,p) = \tau([x]_{q_1}, [y]_{q_2},p) 
\eeq
has a fermionic formula of the form 
\beq
  \rho(x,y,p) 
  = \langle p|
    \exp\left(\sum_{k=1}^\infty\frac{x^k}{k(1-q_1^k)}J_k\right)g
    \exp\left(- \sum_{k=1}^\infty\frac{y^k}{k(1-q_2^k)}J_{-k}\right)
    |p \rangle 
\eeq
and satisfies the equation 
\begin{multline}
  \rho(q_1x,y,p)\rho(x,q_2y,p) - \rho(x,y,p)\rho(q_1x,q_2y,p) \\
  = xy\rho(x,q_2y,p+1)\rho(q_1x,y,p-1). 
\end{multline}
This is another version of the $q$-difference Toda equation, 
which is rather common in the literature 
\cite{MMV94,AHvM98,Takasaki05}. 
\item[3.] 
One can consider the functions 
\beq
\begin{aligned}
  \tilde{\sigma}(x,y,p) 
&= \sigma(x,y,p)\prod_{m,n=0}^\infty(1 - xyq_1^mq_2^n),\\
  \tilde{\rho}(x,y,p) 
&= \rho(x,y,p)\prod_{m,n=0}^\infty(1 - xyq_1^mq_2^n)^{-1} 
\end{aligned}
\eeq
in place of $\sigma(x,y,p)$ and $\rho(x,y,p)$.  
In the usual formulation of the Toda hierarchy, 
these modified tau functions amounts to 
\beq
  \tilde{\tau}(\bsT,\bsTbar,p) 
  = \tau(\bsT,\bsTbar,p)
    \exp\left(- \sum_{k=1}^\infty kT_k\Tbar_k\right), 
\eeq
which is suited for the construction of 
the so called Wronskian solutions of the Toda hierarchy.  
$\tilde{\sigma}(x,y,p)$ and $\tilde{\rho}(x,y,p)$ 
satisfy the $q$-difference equations 
\begin{multline}
  \tilde{\sigma}(q_1x,q_2y,p)\tilde{\sigma}(x,y,p) 
  - (1-xy)\tilde{\sigma}(x,q_2y,p)\tilde{\sigma}(q_1x,y,p) \\
  = xy\tilde{\sigma}(x,y,p+1)\tilde{\sigma}(q_1x,q_2y,p-1) 
\end{multline}
and 
\begin{multline}
  \tilde{\rho}(q_1x,y,p)\tilde{\rho}(x,q_2y,p) 
  - (1-xy)\tilde{\rho}(x,y,p)\tilde{\rho}(q_1x,q_2y,p) \\
  = xy\tilde{\rho}(x,q_2y,p+1)\tilde{\rho}(q_1x,y,p-1). 
\end{multline}
The equation for $\tilde{\rho}(x,y,p)$ coincides 
with the $q$-difference Toda equation of 
Kajiwara and Satsuma \cite{KS91}.  
\end{itemize}

\subsection{$q$-difference equation for $Z(\bst,Q,p;q_1,q_2)$}

As already mentioned in the beginning of this section, 
$Z(\bst,Q,p;q_1,q_2)$ is equal to a special value 
of the function $\sigma(x,y,p)$ defined by 
the fermionic formula (\ref{sigma(x,y,p)-fermion}) 
with $g = Q^{L_0}e^{H(\bst)}$, namely, 
\beq
  Z(\bst,Q,p;q_1,q_2) = \sigma(q_1^{1/2},q_2^{1/2},p).  
\eeq
To derive a $q$-difference equation with respect to $Q$, 
it is more convenient to consider this relation 
in a slightly more general form as follows.  

\begin{lemma}
This special solution $\sigma(x,y,p)$ of the $q$-difference 
Toda equation is related to the partition function 
$Z(\bst,Q,p;q_1,q_2)$ as 
\beq
  \sigma(x,y,p) 
  = (q_1^{-1/2}q_2^{-1/2}xy)^{-p(p+1)/2} 
    Z(\bst,q_1^{-1/2}q_2^{-1/2}xyQ,p;q_1,q_2). 
\label{sigma(x,y,p)-Z(t,Q,p)}
\eeq
\end{lemma}

\proof
Since $L_0$ and $J_k$'s satisfy the well known 
commutation relations 
\beqnn
  [L_0,J_k] = - J_k, 
\eeqnn
we can rewrite the exponential operators 
in (\ref{sigma(x,y,p)-fermion}) as 
\beqnn
\begin{aligned}
  \exp\left(\sum_{k=1}^\infty\frac{x^k}{k(1-q_1^k)}J_k\right) 
 &= (q_1^{-1/2}x)^{-L_0} 
    \exp\left(\sum_{k=1}^\infty\frac{q_1^{1/2}}{k(1-q_1^k)}J_k\right)
    (q_1^{-1/2}x)^{L_0}, \\
  \exp\left(\sum_{k=1}^\infty\frac{y^k}{k(1-q_2^k)}J_{-k}\right) 
 &= (q_2^{-1/2}y)^{L_0} 
    \exp\left(\sum_{k=1}^\infty\frac{q_2^{1/2}}{k(1-q_1^k)}J_{-k}\right)
    (q_2^{-1/2}y)^{-L_0}. 
\end{aligned}
\eeqnn
Consequently, 
\begin{multline*}
  \sigma(x,y,p) 
= \langle p|(q_1^{-1/2}x)^{-L_0} 
  \exp\left(\sum_{k=1}^\infty\frac{q_1^{1/2}}{k(1-q_1^k)}J_k\right)
  (q_1^{-1/2}x)^{L_0}Q^{L_0}e^{H(\bst)}\\
  \mbox{}\times (q_2^{-1/2}y)^{L_0} 
  \exp\left(\sum_{k=1}^\infty\frac{q_2^{1/2}}{k(1-q_1^k)}J_{-k}\right)
  (q_2^{-1/2}y)^{-L_0} |p \rangle. 
\end{multline*}
Since 
\beqnn
\begin{aligned}
  \langle p|(q_1^{-1/2}x)^{-L_0} 
  &= (q_1^{-1/2}x)^{-p(p+1)/2}\langle p|,\\
  (q_2^{-1/2}y)^{-L_0} |p \rangle
  &= (q_2^{-1/2}y)^{-p(p+1)/2}|p \rangle 
\end{aligned}
\eeqnn
and 
\beqnn
  (q_1^{-1/2}x)^{L_0}Q^{L_0}e^{H(\bst)}(q_2^{-1/2}y)^{L_0}
  = (q_1^{-1/2}q_2^{-1/2}xyQ)^{L_0}e^{H(\bst)}, 
\eeqnn
this expression of $\sigma(x,y,p)$ boils down to 
(\ref{sigma(x,y,p)-Z(t,Q,p)}).  
\qed
\bigskip

We can now derive a $q$-difference equation 
for $Z(\bst,Q,;q_1,q_2)$ from the $q$-difference 
2D Toda equation (\ref{q-diff-2DToda}) as follows.  

\begin{theorem}
$Z(\bst,Q,p;q_1,q_2)$ satisfies the $q$-difference equation 
\begin{multline}
  Z(\bst,q_1q_2Q;q_1,q_2)Z(\bst,Q,;q_1,q_2) 
  - Z(\bst,q_1Q,;q_1,q_2)Z(\bst,q_2Q,;q_1,q_2)\\
  = (q_1q_2)^{p+1/2}Z(\bst,Q,p+1;q_1,q_2)Z(\bst,q_1q_2Q,p-1;q_1,q_2). 
\label{q-diff-Z(t,Q,p)}
\end{multline}
\end{theorem}

\proof
Plugging $\sigma(x,y,p)$ of 
(\ref{sigma(x,y,p)-Z(t,Q,p)}) into 
(\ref{q-diff-2DToda}) yields the equation 
\begin{multline*}
  Z(\bst,q_1^{1/2}q_2^{1/2}xyQ,;q_1,q_2)
    Z(\bst,q_1^{-1/2}q_2^{-1/2}xyQ;q_1,q_2)\\
  \mbox{} - Z(\bst,q_1^{1/2}q_2^{-1/2}xyQ,;q_1,q_2)
      Z(\bst,q_1^{-1/2}q_2^{1/2}xyQ;q_1,q_2)\\
  = (q_1q_2)^{p+1/2}Z(\bst,q_1^{-1/2}q_2^{-1/2}xyQ,p+1;q_1,q_2)
      Z(\bst,q_1^{1/2}q_2^{1/2}xyQ,p-1;q_1,q_2). 
\end{multline*}
Upon rescaling $Q$ as $Q \to x^{-1}y^{-1}q_1^{1/2}q_2^{1/2}Q$, 
this equation turns into (\ref{q-diff-Z(t,Q,p)}). 
\qed
\bigskip

It will be instructive to compare this result 
with the reduction process from 
the 2D Toda equation (\ref{2DToda}) 
to the 1D Toda equation 
\beq
  \frac{\rd^2\tau(t,p)}{\rd t^2}\tau(t,p) 
  - \left(\frac{\rd\tau(t,p)}{\rd t}\right)^2 
  = \tau(t,p+1)\tau(t,p-1). 
\label{1DToda}
\eeq
(\ref{1DToda}) is obtained from (\ref{2DToda}) 
by assuming the condition that 
\beqnn
  \tau(x,y,p) = \tau(x+y,p), 
\eeqnn
namely, $\tau(x,y,p)$ be a function of 
$t = x + y$ and $p$.  In the same sense, 
if $\sigma(x,y,p)$ is a function of $t = xy$ and $p$, 
namely, 
\beqnn
  \sigma(x,y,p) = \sigma(xy,p), 
\eeqnn
then the $q$-difference 2D Toda equation 
(\ref{q-diff-2DToda}) reduces to 
the $q$-difference analogue 
\beq
  \sigma(q_1q_2t,p)\sigma(t,p) - \sigma(q_1t,p)\sigma(q_2t,p) 
  = t\sigma(t,p+1)\sigma(q_1q_2t,p-1)  
\label{q-diff-1DToda}
\eeq
of the 1D Toda equation (\ref{1DToda}).  
The $q$-difference equation (\ref{q-diff-Z(t,Q,p)}) 
stems from substantially the same idea.  
The apparent discrepancy of (\ref{q-diff-Z(t,Q,p)}) 
and (\ref{q-diff-1DToda}) is due to 
the prefactor $(q_1^{-1/2}q_2^{-1/2}xy)^{-p(p+1)}$ 
on the right hand side of (\ref{sigma(x,y,p)-Z(t,Q,p)}).

\section{Conclusion}

We have found two kinds of integrable structures 
hidden in the partition function $Z(\bst,Q,p;q_1,q_2)$ 
of the generalized melting crystal model.  
The first integrable structure is the bigraded Toda hierarchy 
that determines the $\bst$-dependence of the partition function. 
This integrable structure emerges when $q_1,q_2$ 
and $q$ satisfy the algebraic relations (\ref{q1-q2-q}) 
(Theorems 1 and 2).  This is a natural generalization 
of the main result of our previous work \cite{NT07}.   
The second integrable structure is a $q$-difference analogue 
of the 1D Toda equation  (Theorem 3).  
The role of time variable therein is played by $Q$, 
and this equation holds without any condition 
on $q_1,q_2$ and $q$

It is easy to see that these results still hold 
if a $GL(\infty)$ element of the form 
\beqnn
  h = \exp\left(\sum_{n=-\infty}^\infty 
      c_n{:}\psi_{-n}\psi^*_n{:}\right)
\eeqnn
is inserted in front of $Q^{L_0}$ (though 
the ordering of $h$, $Q^{L_0}$ and $e^{H(\bst)}$ 
can be changed arbitrarily).  
Of particular interest is the case where 
\beqnn
  h = e^{\beta W_0} \quad 
  \mbox{($\beta$ is a constant)}. 
\eeqnn
This amounts to introducing 
a new potential of the form 
\footnote{Note that each sum
 on the left hand side of this definition 
is divergent.  The right hand side  shows 
its regularized form.}
\begin{multline}
  \sum_{i=1}^\infty (p+\lambda_i-i+1)^2 
    - \sum_{i=1}^\infty (-i+1)^2 \\
  = \sum_{i=1}^\infty ((p+\lambda_i-i+1)^2 - (p-i+1)^2) 
    + \frac{p(p+1)(2p+1)}{6}
\end{multline}
with coupling constant $\beta$. 
The partition function 
\beq
  Z(\bst,\beta,Q,p;q_1,q_2) 
  = \langle p|G_{+}(q_1)e^{\beta W_0}Q^{L_0}e^{H(\bst)}
    G_{-}(q_2)|p\rangle. 
\eeq
thus obtained can be found in 5D supersymmetric 
Yang-Mills theory \cite{MNTT04a,MNTT04b}, 
and  also related to topological strings 
on a class of toric Calabi-Yau threefolds and 
Huwitz numbers of the Riemann sphere \cite{CGMPS06}.

\subsection*{Acknowledgements}

The author is grateful to Toshio Nakatsu 
for collaboration that led to this work.  
This work is partly supported by Grant-in-Aid for 
Scientific Research No. 18340061 and No. 19540179 from 
the Japan Society for the Promotion of Science.


\begin{thebibliography}{99}

\bibitem{ORV03}
A. Okounkov, N. Reshetikhin and C. Vafa,  
Quantum Calabi-Yau and classical crystals, 
in: P. Etingof, V. Retakh and I.M. Singer (eds.), 
{\it The unity of mathematics\/}, 
Progr. Math.  {\bf 244}, Birkh\"auser, 2006, 
pp. 597--618. 
% arXiv:hep-th/0309208

\bibitem{MNTT04a}
T. Maeda, T. Nakatsu, K. Takasaki and T. Tamakoshi, 
Five-dimensional supersymmetric Yang-Mills theories 
and random plane partitions, 
JHEP {\bf 0503}  (2005), paper 056. 
% arXiv:hep-th/0412327. 

\bibitem{MNTT04b}
T. Maeda, T. Nakatsu, K. Takasaki and T. Tamakoshi, 
Free fermion and Seiberg-Witten differential in 
random plane partitions, 
Nucl. Phys. {\bf B715} (2005), 275--303. 
% arXiv:hep-th/0412329

\bibitem{Nekrasov02}
N.A. Nekrasov,
Seiberg-Witten prepotential from instanton counting, 
Adv. Theor. Math. Phys. {\bf 7}  (2004), 831--864. 
% arXiv:hep-th/0206161 

\bibitem{NY05}
H. Nakajima and K. Yoshioka, 
Instanton counting on blowup, 
I. 4-dimensional pure gauge theory, 
Invent. Math {\bf 162} (2005), 313--355. 

\bibitem{NO03}
N. Nekrasov and A. Okounkov,
Seiberg-Witten theory and random partitions, 
in: P. Etingof, V. Retakh and I.M. Singer (eds.), 
{\it The unity of mathematics\/}, 
Progr. Math.  {\bf 244}, Birkh\"auser, 2006, 
pp. 525--296. 
% arXiv:hep-th/0306238. 

\bibitem{OP02a}
A. Okounkov and R. Pandharipande, 
Gromov-Witten theory, Hurwitz theory, and completed cycles, 
Ann. Math. {\bf 163} (2006), 517--560. 
% arXiv:math.AG/0204305

\bibitem{OP02b}
A. Okounkov and R. Pandharipande, 
The equivariant Gromov-Witten theory of $\PP^1$, 
Ann.  Math. {\bf 163} (2006), 561--605. 
% arXiv:math.AG/0207233

\bibitem{OR03}
A. Okounkov and N. Reshetikhin,  
Correlation function of Schur process with application 
to local geometry of a random 3-Dimensional Young diagram, 
J. Amer. Math. Soc. {\bf 16}, (2003), 581--603. 
% arXiv:math.CO/0107056 

\bibitem{NT07}
T. Nakatsu and K. Takasaki, 
Melting crystal, quantum torus and Toda hierarchy, 
Commun. Math. Phys. 285 (2009), 445--468. 
%on-line version DOI: 10.1007/s00220-008-0583-5
%arXiv:0710.5339 [hep-th] 

\bibitem{Takasaki08}
K. Takasaki, Integrable structure of melting crystal 
with external potentials, 
arXiv:0807.4970 [math-ph]. 

\bibitem{UT84}
K. Ueno and K. Takasaki, 
Toda lattice hierarchy, 
Adv. Stud. Pure Math. {\bf 4} (1984), 1--95. 

\bibitem{TT95}
K. Takasaki and T. Takebe, 
Integrable hierarchies and dispersionless limit, 
Rev. Math. Phys. {\bf 7} (1995), 743--808.  

\bibitem{Losev-etal03} 
A. Losev, A. Marshakov and N. Nekrasov, 
Small instantons, little strings and free fermions, 
M. Shifman, A. Vainstein and J. Wheater (eds.), 
Ian Kogan memorial volume, {\it From fields to strings: 
circumnavigating theoretical physics\/}, 
World Scientific, 2005, pp. 581--621. 
% arXiv:hep-th/0302191.  

\bibitem{MN06}
A. Marshakov and N. Nekrasov,  
Extended Seiberg-Witten theory and integrable hierarchy, 
JHEP {\bf 0701}  (2007), paper 104. 
% arXiv:hep-th/0612019. 

\bibitem{Marshakov07}
A. Marshakov, 
On microscopic origin of integrability in Seiberg-Witten theory, 
Theor. Math. Phys. {\bf 154} (2008), 362--384.  
% arXiv:0706.2857 [hep-th]. 

\bibitem{IKS08}
A. Iqbal, C. Koz\c{c}az and K. Shabbir, 
Refined topological vertex, cylindric partitions 
and $U(1)$ adjoint theory, 
arXiv:0803.2260 [hep-th]. 

\bibitem{Carlet06}
G. Carlet, 
The extended bigraded Toda hierarchy, 
J. Phys. A: Math. Gen. {\bf 39} (2006), 9411--9435.  

\bibitem{KS91}
K. Kajiwara and J. Satsuma,  
$q$-difference version of the two-dimensional 
Toda lattice equation, 
J. Phys. Soc. Japan {\bf 60} (1991), 3986--3989. 

\bibitem{MMV94}
A. Mironov, A. Morozov and L. Vinet, 
On a c-number quantum $\tau$-function, 
%%Theor. Math. Phys. {\bf 100} (1995), 890-899. 
Teor. Mat. Fiz. {\bf 100} (1994), 119--131. 

\bibitem{AHvM98}
M. Adler, E. Horozov and P. van Moerbeke, 
The solution to the $q$-KdV equation, 
Phys. Lett. {\bf A242} (1998), 139--151.

\bibitem{Takasaki05}
K. Takasaki, 
$q$-analogue of modified KP hierarchy and 
its quasi-classical limit, 
Lett. Math. Phys. {\bf 72} (2005), 165--181. 

\bibitem{Macdonald-book} 
I.G. Macdonald,
Symmetric functions and Hall polynomials, 
Oxford University Press, 1995.

\bibitem{Bressoud-book}
D.M. Bressoud, 
Proofs and confirmations, 
Cambridge University Press, 1999. 

\bibitem{Gao02}
Y. Gao, 
Fermionic and bosonic representation of 
the extended affine Lie algebra $\mathfrak{gl}_N(\CC_q)$, 
Canad. Math. Bull. {\bf 45} (2002), 623--633. 

\bibitem{JM83}
M. Jimbo and T. Miwa, 
Solitons and infinite dimensional Lie algebras, 
Publ. RIMS, Kyoto Univ., {\bf 19} (1983), 943--1001. 

\bibitem{Takebe91}
T. Takebe, 
Representation theoretical meanings of 
the initial value problem for 
the Toda lattice hierarchy I, 
Lett. Math. Phys. {\bf 21} (1991), 77--84; 
ditto II, Publ. RIMS, Kyoto Univ., 
{\bf 27} (1991), 491--503. 

\bibitem{Teo06}
L.-P. Teo, 
Fay-like identities of the Toda lattice hierarchy 
and its dispersionless limit, 
Rev. Math. Phys. {\bf 18} (2006), 1055--1074.  

\bibitem{Takasaki07}
K. Takasaki, 
Differential Fay identities and auxiliary linear problem 
of integrable hierarchies, 
arXiv:0710.5356 [nlin.SI]. 

\bibitem{CGMPS06} 
N. Caporaso, L. Griguolo, M. Mari\~{n}o, S. Pasquetti 
and D. Seminara, 
Phase transitions, double scaling limit, and 
topological strings, 
Phys. Rev. {\bf D75} (2007), paper 046006. 
%arXiv:hep-th/0606120. 

\end{thebibliography}
\end{document}